\documentclass{aa}  
\usepackage{amsmath}
\usepackage{graphicx}
\usepackage{txfonts}
\usepackage{natbib}
\usepackage{changes}
\bibpunct{(}{)}{;}{a}{}{,}
\usepackage{lipsum}
\usepackage{subcaption}         
\usepackage{lscape}                    
\usepackage{placeins}           
                                
\begin{document}

   \title{Prominence Plasma Parameters Maps Inferred From Lyman $\beta$ and Lyman $\gamma$ Observations and Non-LTE Modelling}


%

   \author{Y. Zhang\inst{1}\fnmsep\thanks{Corresponding author: y.zhang.9@research.gla.ac.uk}
        \and N. Labrosse\inst{1}
        \and T. A. Kucera\inst{2}
        \and S. Parenti\inst{3}
        }

   \institute{SUPA School of Physics and Astronomy, University of Glasgow, Glasgow G12 8QQ UK
   \and Heliophysics Science Division, NASA Goddard Space Flight Center, Greenbelt, MD, 20771 USA
   \and Universit\'e Paris-Saclay, Universit\'e Paris Cit\'e, CEA, CNRS, AIM, 91191, Gif-sur-Yvette, France}

   \date{Received September 30, 20XX}

 
  \abstract
   {The first dedicated observation of an off-limb prominence by the Spectral Imaging of the Coronal Environment (SPICE) instrument took place on April 15, 2023.}
   {We aim to create parameter maps on the prominence region, including temperature, pressure, and column mass, by studying the integrated intensity of the Lyman $\beta$ and Lyman $\gamma$ lines from SPICE data.} 
   {After constraining the altitude and radial velocity in the prominence, we use a 1D non-LTE radiative transfer code to generate 1000 random models and compute the Lyman $\beta$ and Lyman $\gamma$ line profiles. The computed intensities are compared with observed integrated intensities from SPICE. Then, we create models which simultaneously give a reasonable match with the observed intensities in both lines. Unlike previous approaches, our method uses contribution functions to guide the optimisation of temperature and pressure profiles. Our approach enables a physically constrained and consistent match to both spectral lines.}
   {The method in this paper enables us to generate models from pixels on the prominence region and use this information to generate parameter maps. The results obtained have potential for future research.}
   {}

   \keywords{Sun: filaments, prominences --
                Line: formation
               }
   \titlerunning{Parameters maps}
   \maketitle
   \nolinenumbers

\section{Introduction}

Solar prominences are structures of cool and dense plasma suspended within the hot solar corona. The electron temperatures at the cool core are from $7.5 \times 10^{3}$ to $10^{4}\,\mathrm{K}$ and the cool plasma densities are $10^{10} \text{--} 10^{11}~\text{cm}^{-3}$. Solar prominences are pivotal to understanding the dynamics of the solar atmosphere. They are frequently associated with coronal mass ejections (CMEs), which can significantly influence space weather \citep{parenti2014solar}. Investigating the physical conditions within these structures, such as their temperature, pressure, column mass, and velocity, is crucial to advancing our understanding of solar physics and enhancing space weather forecasting.

The determination of these fundamental prominence parameters is primarily based on spectroscopic diagnostics. Temperatures of optically thin plasma, like the prominence-corona transition region (PCTR) and solar jets, can be estimated using Differential Emission Measure (DEM) or filter–ratio techniques with instruments such as SOHO/SUMER or Hinode/XRT \citep[e.g.,][]{parenti2014solar, 2017_Mulay_filter_ratio_jets}. The electron density within prominences core can be inferred from diagnostics such as Thomson scattering measurements by instruments like LASCO \citep[e.g.,][]{susino2019determination}, and the ratio of the \ion{Na}{I} $D_2$ 5890~\AA\ and \ion{Sr}{II} 4078~\AA\ lines\citep{wiehr2016electron, stellmacher2017and}. Accurately interpreting diagnostics within the non-local thermodynamic equilibrium (non-LTE) environment of prominences often requires the use of non-LTE radiative transfer modelling for specific spectral lines \citep[e.g.,][]{labrosse2010physics, parenti2014solar,2015ASSL..415..131L,2015ASSL..415..103H}.  Early work, such as that by \citet{vernazza1981structure}, laid the groundwork for understanding radiative transfer in the solar atmosphere, emphasizing the diagnostic power of hydrogen lines. \citet{gunar2008lyman} demonstrated that a multi-thread model, incorporating small line-of-sight velocities for each thread, can explain the observed asymmetries in Lyman line profiles. \citet{schwartz2015non} employed a statistical comparison between synthetic and observed profiles to constrain key parameters of prominence fine structures, such as column mass and central temperature. The synthetic profiles in these studies were generated using two-dimensional (2D) multi-thread non-LTE models, which assume magnetohydrostatic (MHS) equilibrium and include a PCTR. The PRODOP code, used in this study, is a 1D non-LTE radiative transfer code specifically designed for hydrogen, calcium, helium and magnesium lines in solar prominences and includes consideration of Doppler dimming/brightening effect \citep{gouttebroze2002,labrosse2007effect,levens2019}.

The hydrogen Lyman lines (transitions to the ground state, n=1) serve as powerful diagnostics for prominence plasma. These extreme ultraviolet (EUV) lines are particularly sensitive to plasma properties due to their formation across various optical depths within the prominence. While Lyman $\alpha$ provides a fundamental diagnostic, higher members of the series, such as Lyman $\beta$ and Lyman $\gamma$, offer complementary insights. These higher Lyman lines are formed higher up in the atmosphere compared to Lyman $\alpha$, thereby probing varying thermodynamic conditions ranging from the cool, dense core of the prominence to the hotter PCTR region \citep{vial2007ly, heinzel2001soho}. The ratio of Lyman $\alpha$ to Lyman $\beta$ intensity has been shown to be sensitive to physical and geometrical properties of the fine structure and mass motions within prominences \citep{vial2007ly,gunar2010statistical}. Recent studies have further highlighted the diagnostic potential of multi-wavelength and multi-species observations, such as simultaneous analysis of H$\alpha$ and Mg II lines, to constrain temperature, microturbulence, and optical thickness in quiescent prominences through 1D non-LTE inversions \citep{Barczynski2021,jejvcivc2022non}. Such approaches are also being extended to eruptive prominences observed by instruments like Metis/Solar Orbiter, which combine visible-light and UV diagnostics \citep{heinzel2023first, jejvcivc2022non}. The formation of Lyman $\beta$ and Lyman $\gamma$ lines is sensitive to parameters such as temperature, pressure, and column mass based on  previous research in \cite{zhang2026analysis} and \cite{zhang2026non}, and also \cite{gouttebroze1993hydrogen,heinzel1994theoretical}. By analyzing these lines with non-LTE radiative transfer models, we have the potential to infer variation of these physical properties across prominence structures. 

Furthermore, combined observations in H$\alpha$ and white-light during solar eclipses have enabled the mapping of electron density, effective thickness, and temperature in quiescent prominences, revealing gradients in ionisation structure due to incident radiation \citep{jejvcivc2014multi}. These multi-wavelength diagnostics provide a crucial observational foundation for validating and refining non-LTE models of prominence plasma.

In \cite{zhang2026analysis}, we presented the first comprehensive analysis of the Solar Orbiter/SPICE off-limb prominence observation of 15 April 2023 \citep{2020A&A...642A..14S}, focusing on the spatial and temporal evolution of the Lyman $\beta$ and Lyman $\gamma$ line profiles during the pre-eruptive and eruptive phases. By comparing the integrated intensity and line width distributions in prominence, disk, and coronal regions, \cite{zhang2026analysis} revealed variations indicative of dynamic changes in density, temperature, and optical thickness, and introduced a geometric method to derive radial velocities of eruptive filaments from paired H$\alpha$ images. However, the study also demonstrated that observational diagnostics alone cannot uniquely constrain the underlying physical conditions of the prominence plasma, motivating detailed non-LTE radiative transfer modelling. In \cite{zhang2026non}, we performed detailed 1D non-LTE radiative-transfer modelling (PRODOP) with incident radiation constrained using a SPICE full-disk mosaic, to investigate how key physical parameters control the formation of the Lyman $\beta$ and Lyman $\gamma$ lines. \cite{zhang2026non} introduced the parallel coordinate plot and the elasticity coefficient analysis to explore multi-parameter relations and sensitivities, confirmed the strong relation between H$\alpha$ intensity and emission measure (and revealed an additional temperature dependence of H$\alpha$ intensity), and demonstrated that central temperature, central pressure, column mass and the PCTR temperature gradient are primary factors shaping the emergent Lyman profiles.  By combining the observed intensities with the modelling, \cite{zhang2026non} further provides a technique to constrain the central pressure of prominence parameter with an upper limit $\sim0.48\ \mathrm{dyn\ cm^{-2}}$.  Together, Papers~1 and~2 provide the empirical and theoretical foundation for the analysis presented in this paper.

In the present study we use data from the first dedicated off-limb prominence observation performed by the SPICE instrument aboard Solar Orbiter on April 15, 2023. Our primary objective is to derive key plasma parameters, such as temperature, pressure, and column mass, within the observed prominence region by the Lyman $\beta$ and Lyman $\gamma$ lines observation. By applying a 1D non-LTE radiative transfer code to compare computed intensities with observed SPICE integrated intensities, we aim to construct detailed parameter maps of the prominence. The analysis of the pre-eruptive phase of the prominence will provide valuable insights into the thermodynamic and dynamic properties that set the stage for solar eruptions, thereby contributing to our understanding of solar activity and its implications for space weather.

In Section~\ref{sec:obs}, we describe the instruments used in this work and provide an overview of the prominence observations for the event on April 15, 2023. In Section~\ref{sec:construction}, we introduce the method to construct the prominence model which is close to both observations of Lyman $\beta$ and Lyman $\gamma$ lines. In Section~\ref{sec:map}, we present parameter maps generated by the best coefficient set. In Section~\ref{sec:conclusion}, we summarize our conclusions and discuss the future work we plan to do.

\section{Observations}\label{sec:obs}


The Solar Orbiter mission, developed by the European Space Agency (ESA) \citep{muller2020solar}, follows a highly elliptical orbit around the Sun, with most remote-sensing observations conducted near its perihelion. During the period analyzed in this study, Solar Orbiter was situated at a heliocentric distance of 0.31~AU and approximately in quadrature with Earth.

The Spectral Imaging of the Coronal Environment (SPICE) instrument is an imaging spectrometer designed to operate in the ultraviolet (UV) and extreme ultraviolet (EUV) range. It covers two wavelength bands: 69.6--79.6~nm in the short-wavelength (SW) channel and 96.0--105.8~nm in the long-wavelength (LW) channel \citep{anderson2020solar}. SPICE supports Solar Orbiter’s primary scientific goals by providing comprehensive diagnostics of the physical conditions and elemental composition of the solar atmosphere.

SPICE has contributed directly to Solar Orbiter’s objective of investigating coronal mass ejection (CME) initiation by performing coordinated prominence observations during 10-hour campaigns conducted near perihelion. In the prominence observation on April 15, 2023, we used limb-pointing geometries, allowing the combination of SPICE’s spectroscopic diagnostics with high-resolution imaging from the Extreme Ultraviolet Imager (EUI). We have SPICE observation during 01:03--14:58 UT, 15 April 2023, and use pre-eruption rasters from 07:03 UT to 08:09 UT to analyze the event in this paper. 


\begin{figure}
    \centering
    \includegraphics[width=0.5\textwidth]{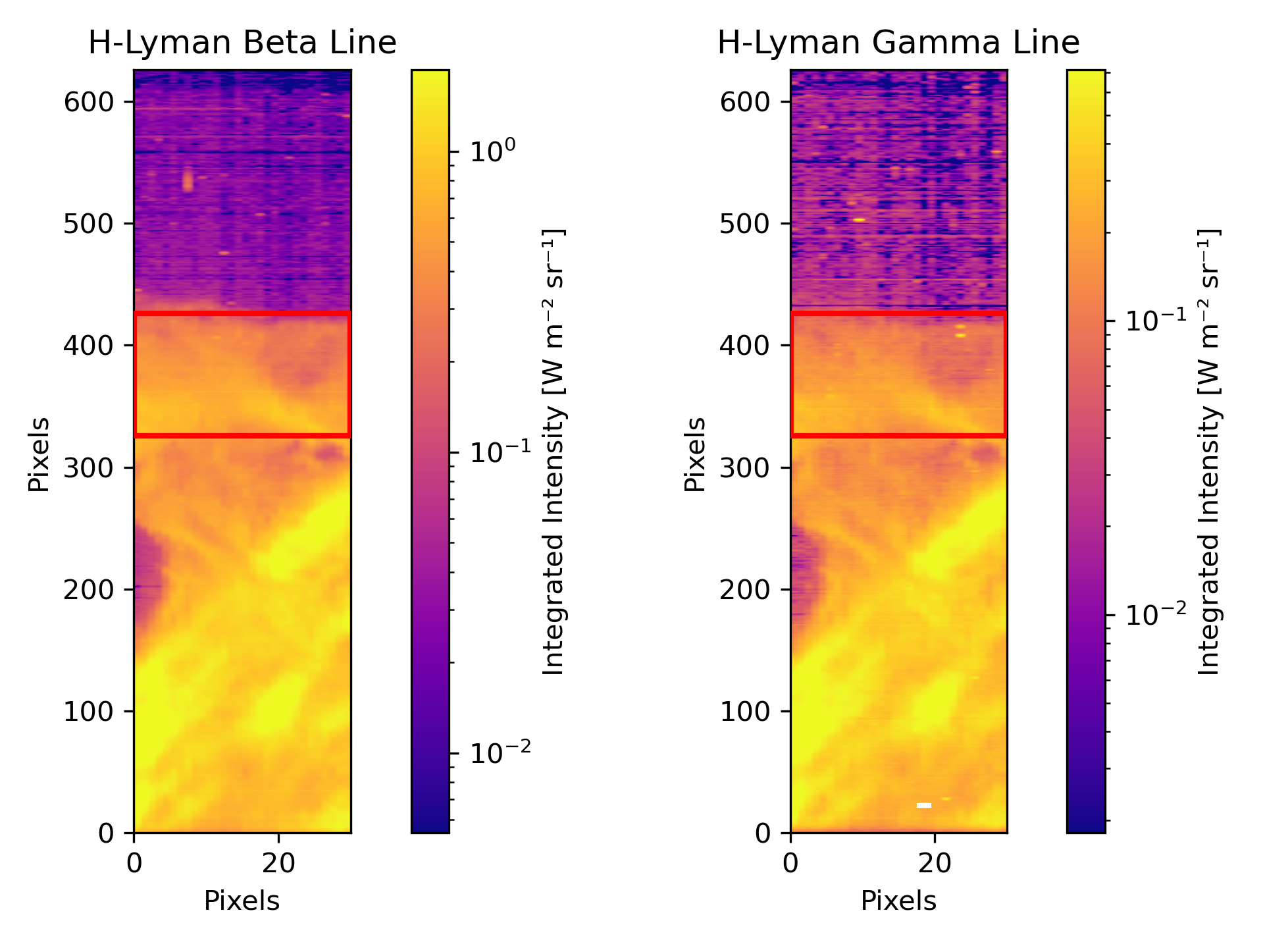}
    \caption{SPICE observation of the integrated intensity of the Lyman $\beta$ and Lyman $\gamma$ lines taken from  07:03:40 - 08:09:23 UT, 15 April 2023. The red rectangle is the prominence region we study. }
    \label{fig:beta_gamma}
\end{figure}

Figure~\ref{fig:beta_gamma} displays the SPICE observation of a prominence, captured in Lyman $\beta$ and Lyman $\gamma$ lines on April 15, 2023, prior to its eruption. we use SPICE data version 5.0 to analyze the event (DOI:10.48326/idoc.medoc.spice.5.0). Analysis of the SPICE data suggests the prominence spanned an altitude range of approximately $20000~\text{km} \sim 60000~\text{km}$ \citep{zhang2026analysis}. 

%


\section{Prominence model atmosphere}\label{sec:construction}

\begin{figure*}
    \centering
    \includegraphics[width=\textwidth]{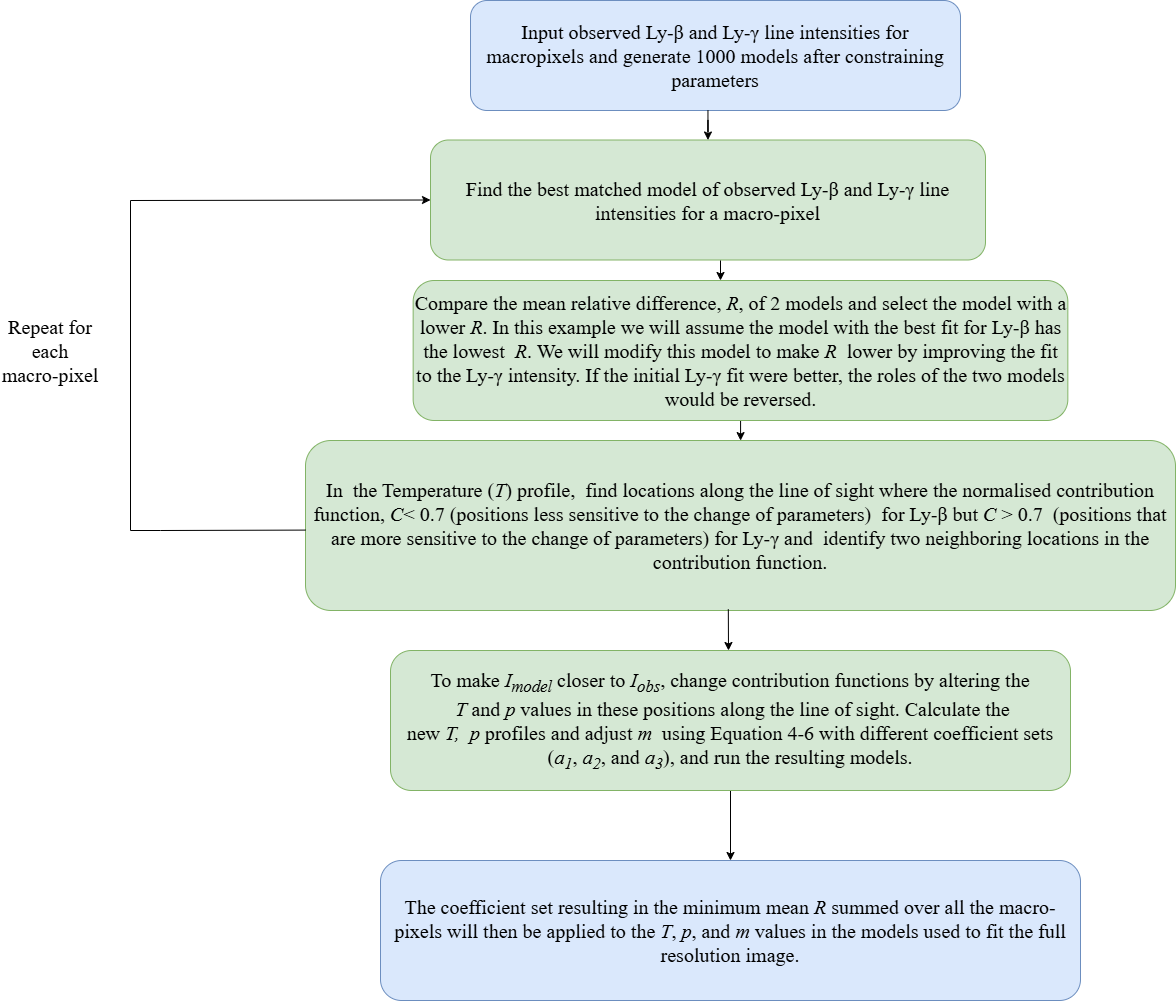}
    \caption{The process to construct initial set of prominence models that are close to observations of the Lyman $\beta$ and Lyman $\gamma$ lines.}
    \label{fig:process}
\end{figure*}

In Figure~\ref{fig:process}, we show the process adopted here to construct prominence models that are close to the observed integrated intensities of both Lyman $\beta$ and the Lyman $\gamma$ line. The general concept is that we identify the best matched models of the observed integrated intensities of both Lyman $\beta$ and the Lyman $\gamma$ line and select the one with lower mean relative difference of Lyman $\beta$ and Lyman $\gamma$ observed intensity. After that, we slightly adjust the model using the other model that has the best match with the other line to make it closer to the observed integrated intensities of both Lyman $\beta$ and the Lyman $\gamma$ line. We will generate new models that can be potentially the best match to observed integrated intensities of both Lyman $\beta$ and the Lyman $\gamma$ line based on different possible coefficient sets. We will calculate new models and find the best coefficient set that can give a lowest mean relative difference for all pixels.

\begin{figure*}
    \centering
    \includegraphics[width=0.9\textwidth]{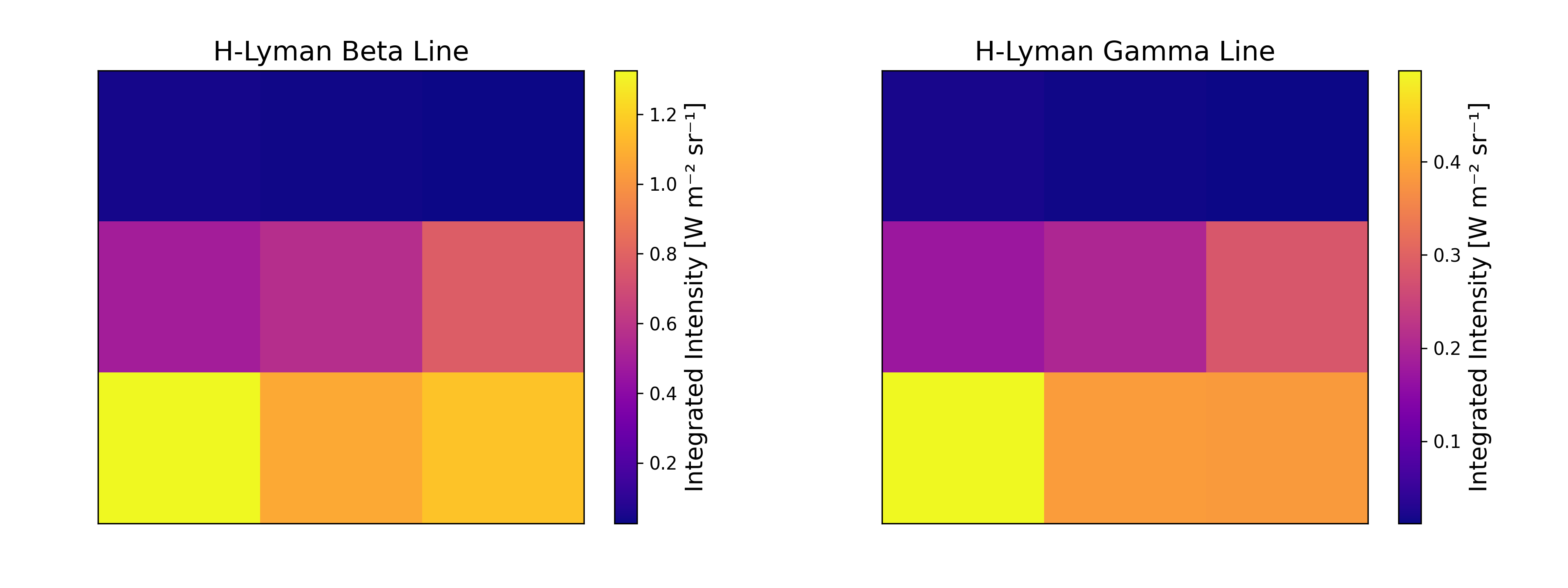}
    \caption{The 9-pixel intensity map of the Lyman $\beta$ and the Lyman $\gamma$ line in the prominence region (the red box region in Figure~\ref{fig:beta_gamma}). For each pixel, the intensity is the mean intensity of that region.}
    \label{fig:9pixel}
\end{figure*}

As a first approach, we divide the intensity map of the Lyman $\beta$ and the Lyman $\gamma$ line in the prominence region (the red box region in Figure~\ref{fig:beta_gamma}) into 9 regions to test our method more efficiently. Figure~\ref{fig:9pixel} shows the mean intensity  for each of these regions. 

We generate 1000 random models based on our estimate of parameters’ ranges in Table~\ref{table:new_parameter_ranges}. These ranges are from \cite{zhang2026non}. We use the Spectral Imaging of the Coronal Environment (SPICE) full-disk mosaic from November 13, 2023 to constrain the incident radiation (the same as \cite{zhang2026non}). Our procedure to obtain the model with the closest intensity to both Lyman $\beta$ and  Lyman $\gamma$ lines is: 

\begin{table}
    \centering
    \caption{\centering The parameter ranges for different variables.}
    \resizebox{\linewidth}{!}{
    \begin{tabular}{|l|l|}
    \hline
    \textbf{Parameter} & \textbf{Range} \\ \hline
    Central Temperature & $5000~\text{K} \sim 15000~\text{K}$ \\
    Surface Temperature & $100000~\text{K}$ \\
    Central Pressure & $0.01~\text{dyn}~\text{cm}^{-2} \sim 0.48~\text{dyn}~\text{cm}^{-2}$ \\
    Surface Pressure & $0.01~\text{dyn}~\text{cm}^{-2}$ \\
    Column Mass & $1.4 \times 10^{-7}~\text{g}~\text{cm}^{-2} \sim 1.4 \times 10^{-5}~\text{g}~\text{cm}^{-2}$ \\
    $\gamma$ (Temperature Gradient) & $2 \sim 10$ \\
    Altitude & $20000~\text{km} \sim 60000~\text{km}$ \\
    Radial Velocity & $0~\text{km}~\text{s}^{-1} \sim 40~\text{km}~\text{s}^{-1}$ \\
    Microturbulent Velocity & $5~\text{km}~\text{s}^{-1}$ \\ \hline
    \end{tabular}
    }
    \label{table:new_parameter_ranges}
\end{table}

\begin{figure*}
    \centering
    \includegraphics[width=\textwidth]{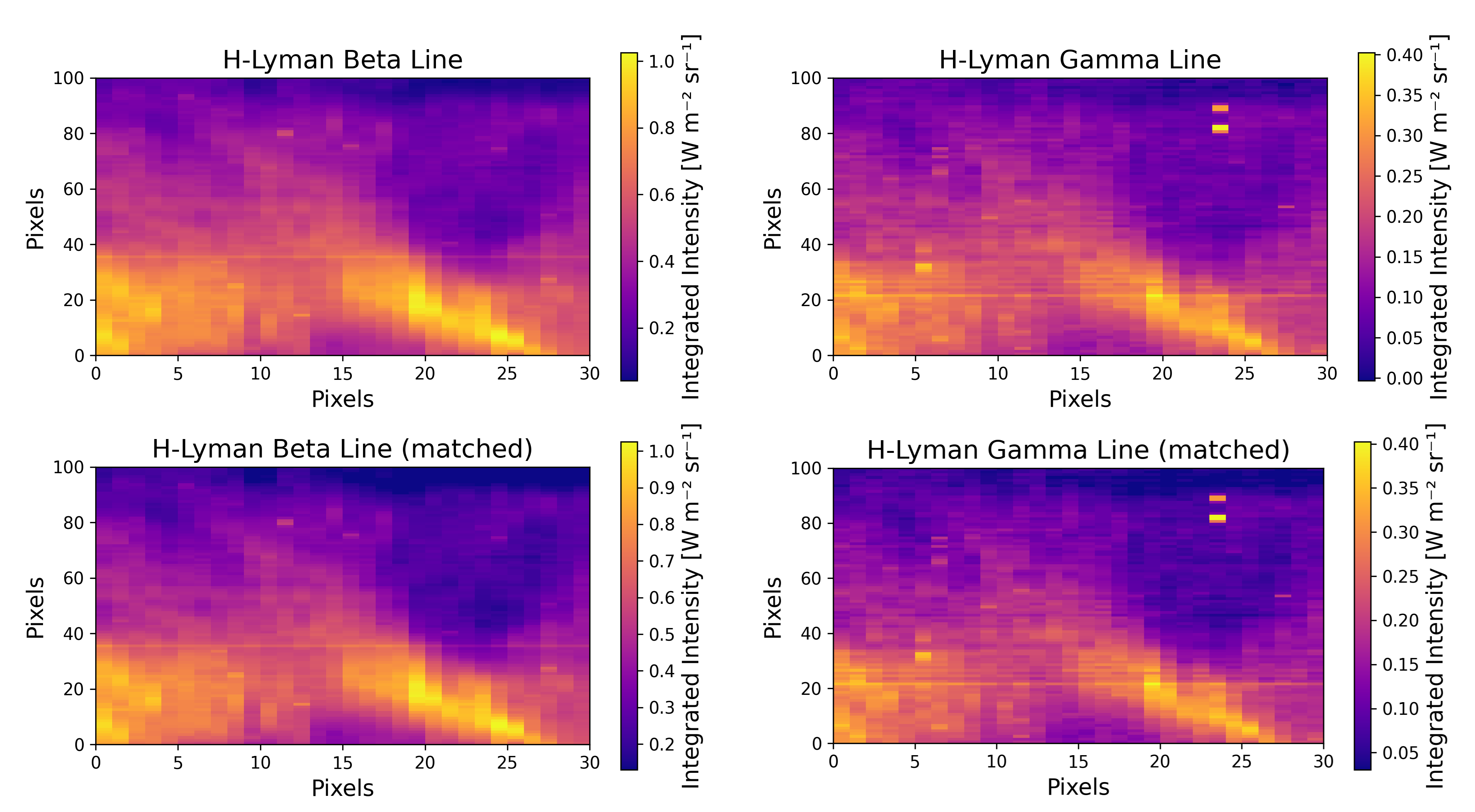}
    \caption{Comparison of best matched intensity map and observation for the red box region in Figure~\ref{fig:beta_gamma}. The upper panels are observational maps and the lower panels are best matched synthetic maps. In the lower panels, pixels in the same position of the Lyman $\beta$ intensity map and the Lyman $\gamma$ intensity map are actually from different models.}
    \label{fig:cal_ob1}
\end{figure*}

(1) For each pixel, we find model A and model B to prepare for the construction of a new model. Model A is the model that has the closest Lyman $\beta$ intensity with observation among the 1000 random models. Model B is the model that has the closest Lyman $\gamma$ intensity with observation among the 1000 random models. The comparison of best matched intensity map and observation for the red box region in Figure~\ref{fig:beta_gamma} is shown in Figure~\ref{fig:cal_ob1}. The calculated intensity maps match quite well with the observation. This is because  we have constrained the range of values for the central pressure based on our results from  \cite{zhang2026non}, based on our observation that the elasticity coefficient between the integrated intensity and the central pressure is high. As a result, the calculated integrated intensities of our 1000 models  fall in the range of integrated intensities in the observation. But in this step, in the lower panels of Figure~\ref{fig:cal_ob1}, pixels in the same position of the Lyman $\beta$ intensity map and the Lyman $\gamma$ intensity map are actually from different models with different plasma parameters. We need to construct new models which can match observations of both Lyman $\beta$ and Lyman $\gamma$ lines simultaneously;

(2) In this step, we will construct new temperature profiles, new pressure profiles, and set column mass for new models. From our previous results in \cite{zhang2026non}, we know that temperature, pressure, and column mass are key parameters of Lyman $\beta$ and Lyman $\gamma$ line formation for the prominence event in April 15, 2023. Therefore, we focus on constructing these three parameters to match the observation. The new temperature and pressure profiles from the prominence surface to the prominence center (the green line in Figure~\ref{fig:9pixelpro}) are obtained by following steps: 
\begin{enumerate}[(i)]
\item we calculate the mean relative difference of Lyman $\beta$ and Lyman $\gamma$ observed intensity ($I_{\beta\_obs}$,  $I_{\gamma\_obs}$), $R$ and calculated intensity for each pixel in model A and model B by Equation~\ref{mrd_single}. 

\begin{equation}
R = \frac{1}{2}\left(\frac{|I_{\beta\_obs}-I_{\beta\_model}|}{I_{\beta\_obs}}+\frac{|I_{\gamma\_obs}-I_{\gamma\_model}|}{I_{\gamma\_obs}}\right)
\label{mrd_single}
\end{equation}

In Equation~\ref{mrd_single}, $I_{\beta\_model}$ and $I_{\gamma\_model}$ are the calculated intensity of these lines. For each pixel, We will identify the smaller $R$ model (model A or model B) and build a new model to make $R$ of this pixel lower in the following steps based on this model;

\item  We retrieve the contribution functions ($C$) for the Lyman $\beta$ line (Model A) and the Lyman $\gamma$ line (Model B) as a function of position from prominence surface to prominence center, and normalize them by their respective maxima. The contribution function describes how a given location inside the prominence contributes to the line intensity and it is an output of the model. The contribution function is expressed as in Equation~\ref{contri} \citep{carlsson1997formation}, where $C$ is the contribution function of the spectral line under consideration,  $S$ is the source function, $\tau$ is the optical depth, and $\chi$ is the absorption coefficient:

\begin{equation}
C = S \exp(-\tau) \, \chi
\label{contri}
\end{equation}

The integrated intensity can be expressed as the integral over position in the slab (z) of the contribution function:

\begin{equation}
I = \int C(z) dz = \int S(z)\, e^{-\tau(z)} \chi(z)\ dz
\label{I_vs_contri}
\end{equation}

The contribution function thus represents the local contribution of each layer to the emergent intensity, and depends upon the source function $S(z)$, the monochromatic opacity per unit length $\chi(z)$, and the exponential attenuation factor $e^{-\tau(z)}$ accounting for absorption along the line of sight.

In Figure~\ref{fig:slab}, we can see that, in the PRODOP code, the prominence is modeled as a 1D slab oriented vertically above the solar surface, incorporating the prominence-corona transition region (PCTR). In Figure~\ref{fig:9pixelpro}, for each panel, the left side of X-axis corresponds to the prominence surface and the right side of X-axis to the prominence center (note that our prominence model atmospheres are symmetrical and thus only one half of the model is shown).

\item For each pixel, we will find positions where we want to modify within the slab using temperature and pressure profiles based on $R$ of Model A and B. For the model with the smaller $R$, one of the lines (Lyman $\beta$ or Lyman $\gamma$) is already well matched but the other line is not, so we avoid modifying positions of new temperature and pressure profiles where changes would significantly affect the intensity of the well matched line. The line which is not well matched is the target of further adjustment, so we modify temperature and pressure profiles where the intensity of this line is most sensitive. At positions with a high value of contribution function, the change of parameters would make the intensity change significantly; while, at positions with low value of contribution function, it would not. Therefore, at locations where the normalized contribution function is below 0.7 (positions that are less sensitive to the change of parameters) for the smaller $R$ model but above 0.7 (positions that are more sensitive to the change of parameters) for the larger $R$ model, we assign the temperature and pressure from the smaller $R$ model, denoted as $T_1$ and $p_1$. We aim to to further reduce $R$ of the model. We then examine the contribution functions for the larger $R$ model. If the larger $R$ model is model A, we check the Lyman $\beta$ line; if the larger $R$ model is model B, we check  the Lyman $\gamma$ line. The temperature and pressure at the location where the contribution function reaches its maximum is denoted as $T_{\text{max}}$ and $p_{\text{max}}$.

\begin{figure}
    \centering
\includegraphics[width=0.5\textwidth]{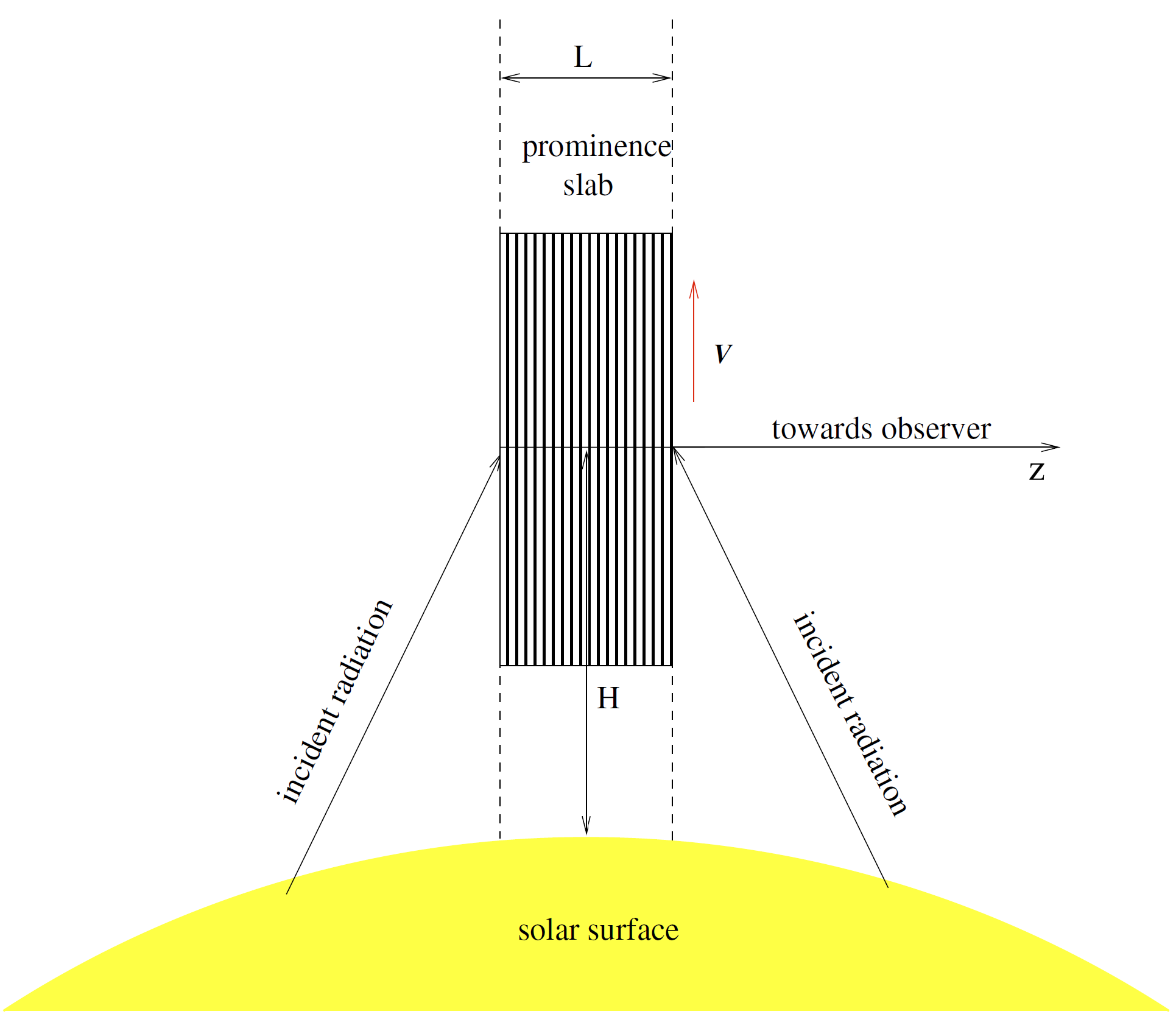}
    \caption{The schematic of 1D prominence slab model in the PRODOP code (Fig.6 in \cite{labrosse2012plasma}).}
    \label{fig:slab}
\end{figure}

\item We will construct new temperature and pressure profiles and set column mass for new models in this step. We check values of two neighbor positions of $T_1$ and $p_1$ in the temperature or pressure profile of the smaller $R$ model. In the smaller $R$ model, we check the calculated intensity of either the Lyman $\beta$ line or Lyman $\gamma$ line, whichever is the worst match. If the calculated intensity is smaller than observed intensity of that line, we need to increase contribution functions, and the neighboring value of $T_1$ (or $p_1)$, which can provide $|T-T_{\text{max}}|$ or $|p-p_{\text{max}}|$ lower than $|T_1-T_{\text{max}}|$ or $|p_1-p_{\text{max}}|$, is denoted as $T_2$ (or $p_2)$; if the calculated intensity is larger than observed intensity of that line, we need to decrease contribution functions, and the neighboring value of $T_1$ (or $p_1$), which can provide $|T-T_{\text{max}}|$ or $|p-p_{\text{max}}|$ larger than $|T_1-T_{\text{max}}|$ or $|p_1-p_{\text{max}}|$, is denoted as $T_2$ (or $p_2)$. 

The temperature and pressure profiles of a new model is calculated by Equations~\ref{linear1} and \ref{linear2}, i.e. a linear combination of $T_1$ and $T_2$ (or $p_1$ and $p_2$). 

\begin{align}
T_{\text{new}} &= T_1 + a_1 \cdot (T_2 - T_1) \label{linear1}\\
p_{\text{new}} &= p_1 + a_2 \cdot (p_2 - p_1)
\label{linear2}
\end{align}

We will repeat the same calculation for all positions that satisfy the condition in the $T$ and $p$ profiles. Since we check neighbor values that are from original $T$ and $p$ profiles, so the order of calculation would not affect results. 

We restrict the coefficients $a_1$ and $a_2$ in Equations~\ref{linear1} and \ref{linear2} to the range $0-1/2$. This choice ensures that the resulting temperature (or pressure) profiles remain closer to $T_1$ (or $p_1$), which belong to the temperature or pressure profile with a smaller $R$. At the same time, the coefficients are not allowed to be negative, because this shifts $T_{\text{new}}$ away from the desired model that has temperature and pressure profiles close to both model A and model B. The smaller $R$ model is already the best matched model of the Lyman $\beta$ line or the Lyman $\gamma$ line. We choose to modify values where the normalised contribution function values (model A, Lyman $\beta$ line; model B, Lyman $\gamma$ line) are below 0.7 for the smaller $R$ model and higher than 0.7 for the larger $R$ model, because we want to reduce the change to the smaller $R$ model and increase the change to the larger $R$ model. If there is no neighboring value which satisfy the condition, then $a_1=a_2=0$.



\begin{figure*}
    \centering
    \includegraphics[width=\textwidth]{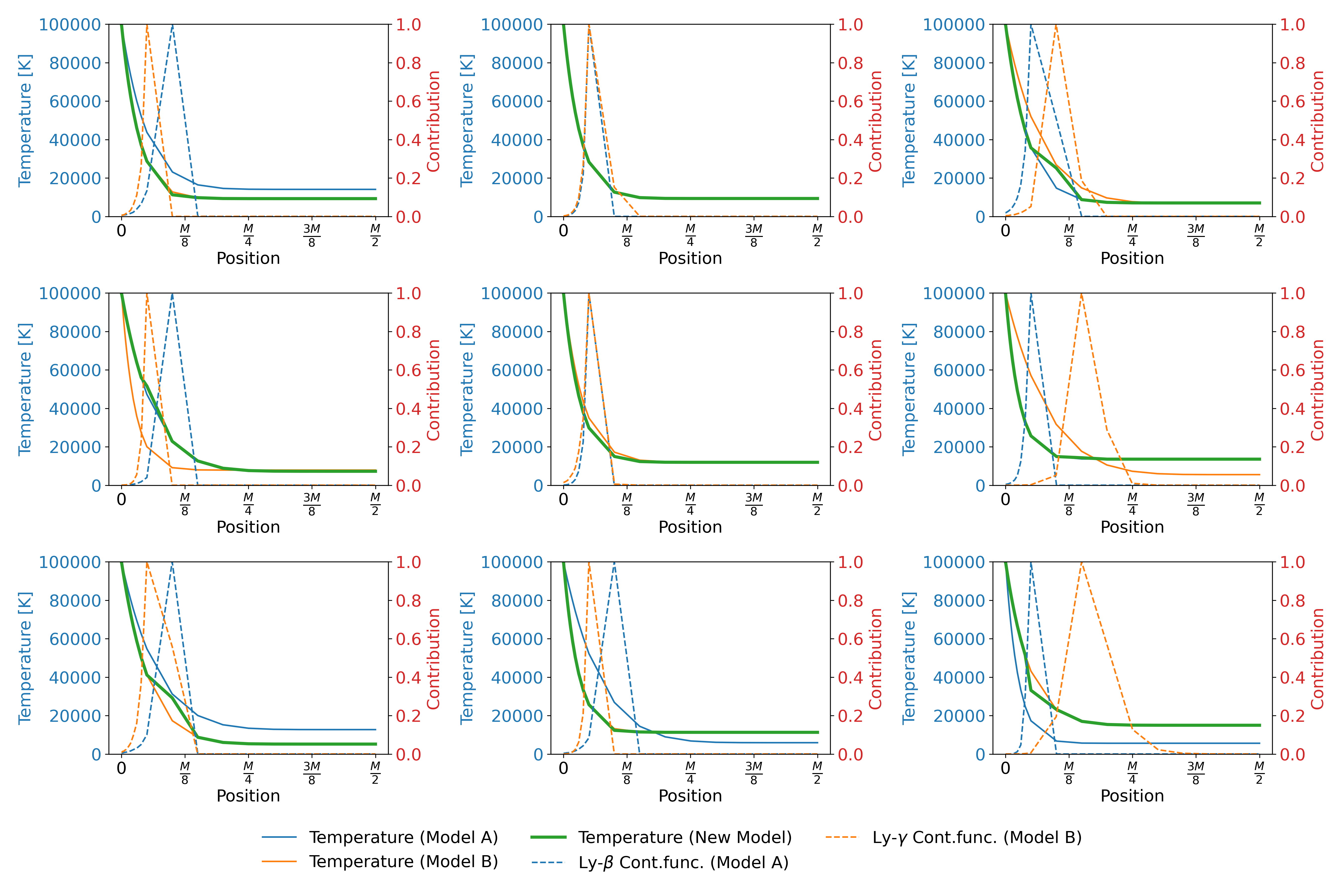}
    \caption{Temperature profiles of Model A and Model B and contribution functions of Model A for the Lyman $\beta$ line and Model B for the Lyman $\gamma$ line. The 9 panels correspond to 9 pixels in Figure~\ref{fig:9pixel}. The solid lines are temperature profiles and the dashed lines are contribution functions. The blue lines are Model A and the orange lines are Model B. The green line is the temperature profile of the new model we construct. $M$ is the column mass of the model, which is along the direction of the line of sight in Figure~\ref{fig:slab}. Contribution functions are normalized by the maximum. In this figure, $a_1=a_2=1/2$.}
    \label{fig:9pixelpro}
\end{figure*}

In this step, the values of $a_1$ and $a_2$ have not been determined. We test them by steps of 1/6: 0, 1/6, 1/3, 1/2. We calculate a new value for the column mass of the new model using Eq.~\ref{linear3}, as column mass is also an important parameter for the formation of both Lyman $\beta$ and Lyman $\gamma$ \citep{zhang2026non}. In Eq.~\ref{linear3}, $m_1$ is the column mass of the smaller $R$ model and $m_2$ is the column mass of the other model. $a_3$ varies from -1/2 to 1/2. We do not consider values larger than 1/2 or lower than -1/2 because this will make either the Lyman $\beta$ and Lyman $\gamma$ line intensity change a lot from original intensity of a model. We also test different values of $a_3$ by an interval of 1/6: -1/2, -1/3, -1/6, 0, 1/6, 1/3, 1/2. In practice, we do not need to test all values since the new column mass 
$m_{\text{new}}$ would be below 0 with some values of $a_3$ below 0. We discard these values directly. We use interval 1/6 after considering both the running time of the code and accuracy of the result and find a balance. 
The result would be more accurate when the interval is smaller, but the running time would be shorter when the interval is larger;

Column mass does not directly affect the radiative transitions process. Instead, it primarily controls the overall density scale and geometrical thickness of the model atmosphere. For this reason, varying the column mass within a wider range does not immediately force the local plasma conditions to deviate strongly from those of Models A and B. By contrast, the temperature and pressure have a much more direct influence on the formation of the Lyman lines; therefore the coefficients $a_1$ and $a_2$ must be restricted to a narrower range to avoid a large change in integrated intensity of both Lyman $\beta$ and Lyman $\gamma$ lines. This is why the allowed interval for $a_3$ in Eq.~\ref{linear3} is chosen to be broader than that for $a_1$ and $a_2$.

\begin{align}
m_{\text{new}} &= m_1 + a_3 \cdot (m_2 - m_1) \label{linear3}
\end{align}

\end{enumerate}

\begin{figure}
    \centering
    \includegraphics[width=0.5\textwidth]{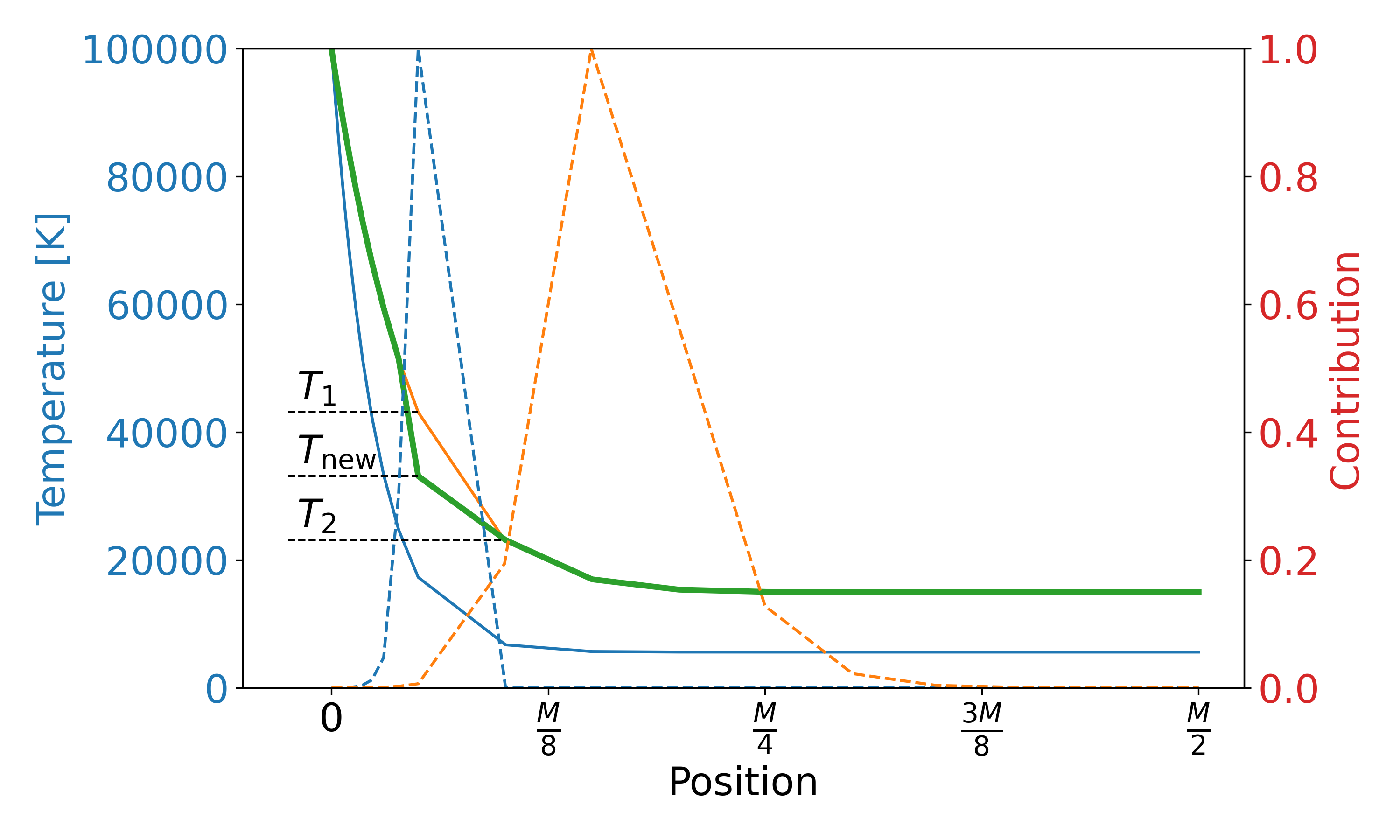}
    \caption{The lower-right panel of Figure~\ref{fig:9pixelpro} showing the details of the changes in the temperature profile.}
    \label{fig:example}
\end{figure}

In Figure~\ref{fig:example}, we can take the lower-right macro-pixel in Figure~\ref{fig:9pixelpro} as an example. In this macro-pixel, after comparing $R$ of model A and model B, we find that model B is the model with smaller $R$, so we proceed using this model. Then, we will try to construct new temperature and pressure profiles to make $R$ lower. We will find locations where the normalized contribution function is below 0.7  for model B but above 0.7 for model A. As shown in Figure~\ref{fig:example}, there is only one position which satisfies the condition in this macro-pixel (For other macro-pixels, positions which satisfies the condition can be more or can be zero). We assign the temperature from model B in that position, denoted as $T_1$. From this figure, we can see that $T_{max}$ of model A is lower than $T_1$. In model B, the Lyman $\gamma$ line matches the observations well while Lyman $\beta$ does not. We also check that the calculated intensity of Lyman $\beta$ is less than the observed intensity of Lyman $\beta$. We need to make $T_1$ closer to $T_{max}$ to increase contribution function in that position so as to increase the calculated intensity of Lyman $\beta$. So, from two neighbor positions of $T_1$, we choose to the right as $T_2$. We obtain $T_{new}$ by Equation~\ref{linear1}. In the example, $a_1=a_2=1/2$, this is just one of the values we will test. For the pressure profile, we will also calculate the new one by the same method.

(3) In this step, we will find the best coefficient set by running models constructed in step 2. We calculate the line intensity for models generated by Step 2. These models are combination of different values of $a_1$, $a_2$ and $a_3$. We compare the calculated intensity with the observed intensity and calculation mean relative difference for all 9 pixels by Equation~\ref{mrd_all}. In Equation~\ref{mrd_all}, $N_{Pix}$ is the number of pixels. We will find the model with the least mean relative difference with the observed intensity. This is the model that we ultimately select as it simultaneously has the closest Lyman $\beta$ and Lyman $\gamma$ intensity with observation.

\begin{equation}
R_{avg} = \frac{1}{{2N_{Pix}}}\sum \left(\frac{|I_{\beta\_obs}-I_{\beta\_model}|}{I_{\beta\_obs}}+\frac{|I_{\gamma\_obs}-I_{\gamma\_model}|}{I_{\gamma\_obs}}\right)
\label{mrd_all}
\end{equation}

For the red box region in Figure~\ref{fig:beta_gamma}, the final results of our method are: 

\begin{align}
a_1 &= 0,  a_2=0,  a_3=0 \notag
\end{align}

This coefficient set gives a $R_{avg}$ of 5~\% between models and observations for the 9-pixel intensity map. The running time of the code is about 4~h (1576 models have been calculated). The coefficient set seems to show that the construction of new temperature and pressure profiles is not necessary. We think there are reasons for this result. One is that the PRODOP code can already match the Lyman $\beta$ and Lyman $\gamma$ intensities with observations quite well simultaneously. Another reason is that the Lyman $\beta$ and Lyman $\gamma$ intensities are actually quite sensitive to the change of parameters like temperature, pressure and column mass. The interval of 1/6 for each coefficient may be too large: a smaller interval could lead to a slightly different coefficient set. Nevertheless, the $R_{avg}$ of 5~\% between models and observations is good enough for our purpose, and the selected models are adequate. We can conclude that in this particular context, the selected coefficient set is the best one. The steps that we have taken to select these coefficients will be important when we extend the method to other lines that have different line formation mechanisms.

\begin{figure*}
    \centering
    \includegraphics[width=\textwidth]{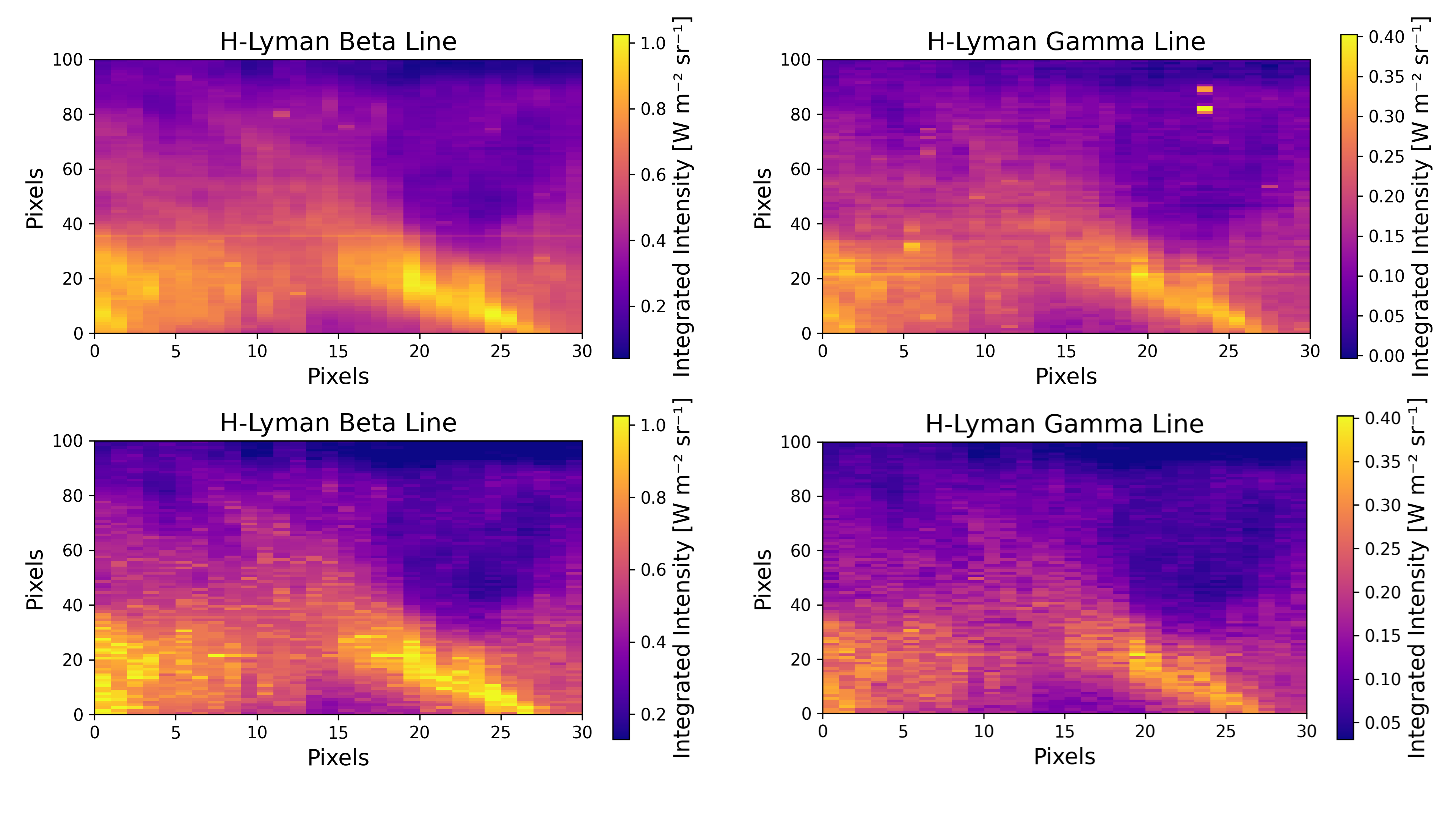}
    \caption{The comparison of calculated intensity by the best coefficient set and observation for the red box region in Figure~\ref{fig:beta_gamma}. The upper panels are observational maps and the lower panels are synthetic maps. In the lower panels, pixels in the same position of the Lyman $\beta$ intensity map and the Lyman $\gamma$ intensity map are from same models.}
    \label{fig:cal_ob2}
\end{figure*}

We also calculate the mean relative difference between the calculated intensity and original intensity map by the same coefficient set. The result is 8~\%, which is quite close to the result of the 9-pixel intensity map. 
The 9-pixel intensity map is a good approach to find the best coefficient set, as the running time prevents the use of the original intensity map to compute the coefficient set. The comparison of calculated intensity by the best coefficient set and observation for the red box region in Figure~\ref{fig:beta_gamma} is shown in Figure~\ref{fig:cal_ob2}. From this figure, we can see that the calculated integrated intensity maps are quite close to observations for both lines.


\begin{figure*}
    \centering
    \makebox[\linewidth]{%
        \includegraphics[width=1.1\textwidth]{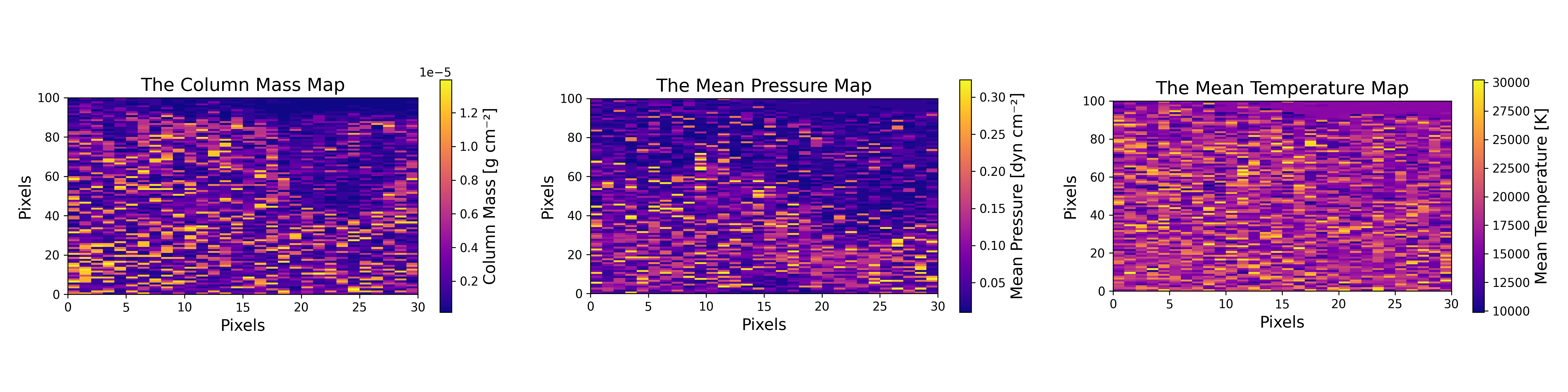}
    }
    \caption{Parameter maps for the most important parameters in the Lyman $\beta$ and Lyman $\gamma$ line formation.}
    \label{fig:parameter_map}
\end{figure*}

\section{Parameter maps}\label{sec:map}

We obtain the plasma parameters for the models generated by the best coefficient set. For the most important parameters in the Lyman $\beta$ and Lyman $\gamma$ line formation, we present parameter maps in Figure~\ref{fig:parameter_map}. 
The standard deviations of the column mass map, the mean pressure map, and the mean temperature map are $3.4 \times 10^{-6}~\mathrm{g\,cm^{-2}}$, $0.07~\mathrm{dyn\,cm^{-2}}$, and $3987~\mathrm{K}$, respectively. These values correspond to $95\%$, $90\%$, and $22\%$ of their respective mean values. The values of integrated intensity of the Lyman $\beta$ and Lyman $\gamma$ line in lower left quadrant are more \added{uniform} than other regions, we can use the correspond percentages of standard deviations to their respective mean values in this region to estimate uncertainties of derived parameters. The uncertainties of the column mass map, the mean pressure map, and the mean temperature map are $86\%$, $72\%$, and $22\%$. In Table~\ref{tab:parameter_maps}, we summary the information of parameter maps.

\begin{table*}[htbp]
\centering
\caption{Summary of parameter maps provided by the best coefficient set: ranges, standard deviations, and estimated uncertainties.}
\label{tab:parameter_maps}
\begin{tabular}{lccc}
\hline
Parameter & Range & Standard Deviation (std) & Uncertainty (\%) \\
\hline
Column Mass [$\mathrm{g\,cm^{-2}}$] & $1.6\times10^{-7} \sim 1.4\times10^{-5}$ & $3.4\times10^{-6}$ & 86\% \\
Mean Pressure [$\mathrm{dyn\,cm^{-2}}$] & $0.01 \sim 0.32$ & 0.07 & 72\% \\
Mean Temperature [K] & $9885 \sim 30238$ & 3987 & 22\% \\
\hline
\end{tabular}
\end{table*}

The variation from pixel to pixel in parameter maps is due to the uncertainty associated with the degeneracy of the solutions. Different sets of parameters can result in the same integrated intensities in some lines. For instance, high pressure and low temperature, or low pressure and high temperature, can lead to the same integrated intensity of Lyman $\beta$ line. If we focus on a region that has similar Lyman $\beta$ intensity in Figure~\ref{fig:cal_ob2} and  in Figure~\ref{fig:parameter_map}, we can verify this. The parameter values are within the ranges quoted by \cite{parenti2014solar}. These values are also consistent with the typical plasma parameters for quiescent prominences summarized in Table~1 by \cite{labrosse2010physics}. We can compare structures in Figure~\ref{fig:cal_ob2} with Figure~\ref{fig:parameter_map}. For the column mass map and the mean pressure map, we can see that the low column mass regions and the low mean pressure parts are roughly low intensity parts in Figure~\ref{fig:cal_ob2} and the high column mass parts and high mean pressure parts are roughly high intensity parts. 

Equations~\ref{meanpre} and \ref{meantem} give the expressions for the mean pressure and mean temperature, respectively, shown in Figure~\ref{fig:parameter_map}. The mean pressure is the average pressure distributed over a unit column mass, and the mean temperature is the average temperature distributed over a unit column mass \citep{anzer1999energy}. $T_{\text {cen}}$ is the central temperature and $T_{\text {tr}}$ is the boundary transition-region temperature. $\gamma$ is the steepness of the temperature gradient in the PCTR. $p_0$ is the pressure at the outer boundary and $p_{\text {cen}}=p_c+p_0$. $p_c$ is the difference between the pressure at the center and at the outer boundary. The column mass $m$ ranges from $m=0$ to $m=M$ (from one surface to the opposite surface). 

\begin{equation}
\overline{p} = \frac{1}{M} \int_{0}^{M} 4p_c \frac{m}{M} \left(1 - \frac{m}{M}\right) + p_{\mathrm{tr}} ~ dm \\
 = \frac{2p_{\mathrm{cen}} + p_{\mathrm{tr}}}{3}
\label{meanpre}
\end{equation}

\begin{equation}
\begin{aligned}
\overline{T}&=\frac{2}{M}\int_0^{\frac{M}{2}}T_\mathrm{cen}+(T_\mathrm{tr}-T_\mathrm{cen})\left(1-4\frac{m}{M}\left(1-\frac{m}{M}\right)\right)^\gamma dm \\
&= \frac{2\gamma T_\mathrm{cen} + T_\mathrm{tr}}{2\gamma + 1}
\label{meantem}
\end{aligned}
\end{equation}

We choose to study mean pressure and mean temperature because they present a useful overall picture of the prominence. In \cite{zhang2026non}, we study the effect of central temperature and central pressure with integrated intensity. Since we fix the surface temperature and surface pressure, the mean pressure and mean temperature are directly related to central temperature and central pressure. Therefore, they have similar relationships with integrated intensity. For the mean temperature map, the structures are not so obvious, and we can only distinguish the upper-right part. There are two reasons for this. One is that the Lyman $\beta$ and Lyman $\gamma$ line intensities are not so sensitive to mean temperature compared with column mass and mean pressure \citep{zhang2026non}. The difference of the Lyman $\beta$ and Lyman $\gamma$ line intensities in the red box region is not large enough to show the difference in mean temperature in this region. This is consistent with the understanding from non-LTE prominence modeling that different spectral lines exhibit varying sensitivities to specific plasma parameters \citep{labrosse2010physics}. To obtain parameter maps with more detailed structures, we need to use spectral lines whose intensities have larger correlation coefficients and  elasticity coefficients with each parameter. This will be one of our future works.

\begin{figure*}
    \centering
    \makebox[\linewidth]{%
        \includegraphics[width=\textwidth]{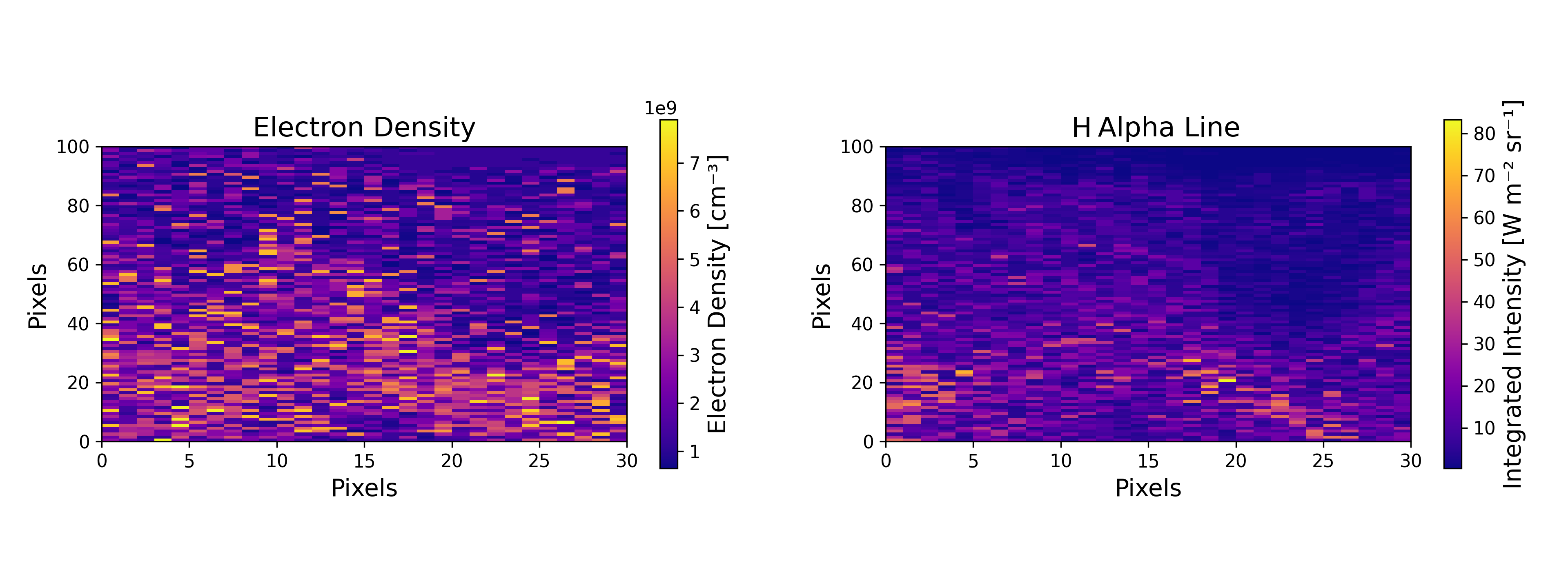}
    }
    \caption{The electron density map and H$\alpha$ intensity map as examples of other maps we can produce.}
    \label{fig:map_example}
\end{figure*}

The maps presented in Figure~\ref{fig:parameter_map} help us to understand the distribution of the parameters that were shown to have the most effect on the calculated Lyman $\beta$ and Lyman $\gamma$ intensities. The non-LTE calculations allow us to study the distribution of other relevant quantities.
For example, in the left panel of Figure~\ref{fig:map_example}, we can see that the electron density map shows similar structures as in the Lyman $\beta$ and Lyman $\gamma$ line observations within the expected range for electron density in a prominence. Each value of the electron density in the map is the mean value along the slab in the direction of light of sight in Figure~\ref{fig:slab}.
We can also produce intensity maps for other lines. In the right panel of Figure~\ref{fig:map_example}, we show the computed H$\alpha$ intensity map, which presents structures very similar to those seen in the Lyman $\beta$ and Lyman $\gamma$ line observations. The visibility of H$\alpha$ intensity map is good compared with other parameter maps, and the structure is quite close to the synthetic Lyman $\beta$ and Lyman $\gamma$ intensity maps with the best coefficient set in Figure~\ref{fig:cal_ob2}. The calculated H$\alpha$ intensities are not large, e.g. compared with another observation in \cite{labrosse2022first}; the prominence discussed here would likely appear as a  faint prominence. 
This allows us to generate synthetic maps for various observables (intensity, width) for spectral lines that aren't available in the observation. \added{ If a spacecraft has a good angle to observe some phenomenons in the sun but it lacks some spectral lines to do analysis or compare with other observations, this method will be useful.}

\section{Conclusion}\label{sec:conclusion}

In this study, we developed a novel modeling approach to derive the plasma parameters of solar prominences based on SPICE observations from the Solar Orbiter mission. Using the first dedicated off-limb prominence raster obtained on 15 April 2023, we devised a model-constructing tool that combines the diagnostic potential of the Lyman $\beta$ and Lyman $\gamma$ lines. By comparing the observed and synthetic intensities through a random-model sampling and refinement process, we achieved a mean relative difference of approximately 8\% between the modeled and observed intensity maps. This relative difference is acceptable in view of the uncertainties inherent to observations (particularly the radiometric calibration) and to the models (e.g. incident radiation). This demonstrates that the proposed method can reliably reproduce the observed radiative properties of the prominence within reasonable computational time.

The parameter maps derived from the best-fit models reveal the spatial distributions of temperature, pressure, column mass, and other key quantities governing the formation of the Lyman lines. Among them, the column mass and pressure structures exhibit strong correspondence with the observed intensity morphology, implying that local variations in mass and gas pressure play dominant roles in shaping the radiative emission pattern of the prominence. The method thus provides a quantitative bridge between spectroscopic diagnostics and plasma modeling, enabling the visualization of internal prominence structures that are otherwise difficult to infer directly from observation.

Our approach can be extended to include additional spectral lines or line ratios, improving the sensitivity to specific physical parameters. Future work will optimize the selection of diagnostic lines to enhance spatial detail in parameter maps. This methodology opens a pathway toward comprehensive multi-line prominence diagnostics using SPICE, IRIS and future EUV spectroscopic datasets.

\section*{Acknowledgments}
We would like to thank the teams involved in the development and operation of the SPICE instruments and data release. We are also grateful to the SunPy team for providing the tools used in our data analysis. The authors thank Lyndsay Fletcher for her comments on this paper.
YZ is supported by the China Scholarship Council (No. 202206120056). NL acknowledges support from UK Research and Innovation's Science and Technology Facilities Council under grant award numbers ST/T000422/1 and ST/X000990/1. TAK was supported by Solar Orbiter/SPICE funding to NASA Goddard Space Flight Center.

Solar Orbiter is a space mission of international collaboration between ESA and NASA, operated by ESA. The development of SPICE has been funded by ESA member states and ESA. It was built and is operated by a multi-national consortium of research institutes supported by their respective funding agencies: STFC RAL (UKSA, hardware lead), IAS (CNES, operations lead), GSFC (NASA), MPS (DLR), PMOD/WRC (Swiss Space Office), SwRI (NASA), UiO (Norwegian Space Agency).
 Solar Orbiter is a space mission of international collaboration between ESA and NASA, operated by ESA. The EUI instrument was built by CSL, IAS, MPS, MSSL/UCL, PMOD/WRC, ROB, LCF/IO with funding from the Belgian Federal Science Policy Office (BELSPO/PRODEX PEA 4000112292 and 4000134088); the Centre National d’Etudes Spatiales (CNES); the UK Space Agency (UKSA); the Bundesministerium für Wirtschaft und Energie (BMWi) through the Deutsches Zentrum für Luft- und Raumfahrt (DLR); and the Swiss Space Office (SSO).

%
\bibliographystyle{aa}
\bibliography{sample701}







   
  




\end{document}